# Password-Based Authentication and The Experiences of End Users

Assumpta Ezugwu[1], Elochukwu Ukwandu[2,*], Celestine Ugwu[1], Modesta Ezema[1], Comfort Olebara[3], Juliana Ndunagu[4], Lizzy Ofusori[5], Uchenna Ome[1]

[1]Department of Computer Science, Faculty of Physical Science, University of Nigeria, Nsukka, Enugu State, Nigeria
[2]Department of Applied Computing and Engineering, Cardiff School of Technologies, Cardiff Metropolitan University, Wales, United Kingdom
[3]Department of Computer Science, Faculty of Physical Science, Imo State University, Owerri, Imo State, Nigeria
[4]Department of Computer Science, Faculty of Sciences, National Open University of Nigeria, Abuja, Nigeria
[5]School of Management, Information Technology and Governance, University of KwaZulu-Natal, WestVille Campus, Durban, South Africa

**Author's email addresses**: ({assumpta.ezugwu, celestine.ugwu, modesta.ezema, uchenna.ome}@unn.edu.ng), eaukwandu@cardiffmet.ac.uk, chy_prime@yahoo.com, jndunagu@noun.edu.ng, lizzyofusori@yahoo.co.uk
*__Correspondent Author__: Elochukwu Ukwandu (eaukwandu@cardiffmet.ac.uk)

**Abstract**
Passwords are used majorly for end-user authentication in information and communication technology (ICT) systems due to its perceived ease of use. The use for end-user authentication extends through mobile, computers and network-based products and services. But with the attendant issues relating to password hacks, leakages, and theft largely due to weak, reuse and poor password habits of end-users, the call for passwordless authentication as alternative intensifies. All the same, there are missing knowledge of whether these password-based experiences are associated with societal economic status, educational qualification of citizens, their age and gender, technological advancements, and depth of penetration. In line with the above, understanding the experience of end-users in developing economy to ascertain their password-based experience has become of interest to the researchers. This paper aims at measuring the experience of staff and students in University communities within southeastern Nigeria on password-based authentication systems. These communities have population whose age brackets are majorly within the ages of 16 and 60 years; have people with requisite educational qualifications ranging from Diploma to Doctorate degrees and constitutes good number of ICT tools consumers. The survey had 291 respondents, and collected data about age, educational qualifications, and gender from these respondents. It also collected information about their password experience in social media network, online shopping, electronic health care services, and internet banking. Our analysis using SPSS and report by means of descriptive statistics, frequency distribution, and Chi-Square tests showed that account compromise in the geographical area is not common with the respondents reporting good experience with passwords usage. Furthermore, this experience is not in any way related to their age (under 60), and educational qualification. Our experiment did not measure the entropy of end-users' passwords, their password hygiene culture and so cannot relate this experience with the strengths of their passwords nor that of their password hygiene culture. The outcome and recommendations of this research will help inform policy and research direction towards password hygiene culture, management, and the potentials or otherwise of passwordless authentication systems in developing economies.

## 1. Introduction
Good number of literatures exists on passwords such as that of Renaud, Otondo, & Warkentin, (2019) on effect of endowment on password strength; AlSabah, Oligeri, & Riley, (2018) on culture and password, and Furnell (2022) on assessing website password practices. Majority of these have shown that passwords, especially text-based are the most popular used authentication method for end-users of information system in computers and network-based products and services (Shay *et al*., 2010). However, due to unhygienic practices such as choosing weak passwords and insecure management of passwords, they are highly vulnerable to exploitation by both internal and external threat actors. In the other hand, the attraction to the use of passwords as argued by Sharma *et al.,* (2010) lies mostly in its simplicity, practicality, ease of use and low cost rather than the security. But are passwords simple to use, of low cost, easy to use and insecure? Authors of this paper wants to submit that the simplicity of passwords use could depend on so many factors such as age, health condition of the user, number of user's application that requires password protections as well as use of password utilities like password managers. Older generations, people living with dementia, end users with multiple online accounts will have challenges managing multiple passwords. Having multiple passwords can be complicated, prone to have issues related to remembering multiple passwords, temptation of reuse of single login credential on multiple accounts and so on.

The security of assets secured using passwords largely depends on the ability of the user to cultivate good password habits such as regular change of passwords, non-reuse of login credentials on multiple systems, and use of strong passwords derived by using multiple characters with special symbols and numbers as well as the length of it. All the same, cultivating some of these habits have been made easier using password managers. Although several studies have explored consumers experiences with password-based authentication system in developed economies (Butler and Butler, 2015; Bilgihan *et al*., 2016; Morrison *et al*., 2021) and the call and rollout of passwordless authentication use cases as alternative intensifies (Jakkal, 2021). For instance, in 2021, Microsoft posits that the development and roll-out of passwordless authentication system is the future of user access management in computer-based systems (Jakkal, 2021). This involves any method that helps to identify a user without the use of password. However, there is a paucity of research focusing on measuring password experience of end-users in developing economies such as Nigeria and relating same with the existing knowledge. This represents a gap in the literature and gives an opportunity for this study to address.

Recent studies by Ugwu *et al*., (2022) in Nigeria suggest that age and educational qualification have no effect on personal cyber hygiene culture of information and communication technologies (ICTs) users. Further studies by Ugwu *et al*., (2023) in Nigeria on the relationship between cyber hygiene culture of Internet users with gender, employment status and academic discipline did not establish any significant relationship between these dependent and independent variables. These studies by Ugwu *et al.;* (2022) and Ugwu *et al.;* (2023) having been done in Nigeria provided more basis for further study – a study that will establish the likelihood or otherwise of a relationship between cyber hygiene culture and password use experience in developing economies. In line with these, this study focuses on finding:

1. What are the end-user experiences in password-based authentication in Southeastern Nigeria?
2. Does age have impact on end-user experience in password-based authentication of this populace?
3. Does gender have impact on end-user experience in password-based authentication of this populace?
4. Does educational level have impact on end-user experience in password-based authentication of this populace?

To answer these questions, descriptive statistics will be deployed, and hypotheses will be raised and tested using appropriate statistical tests.

The experience of end-users of ICT in this instance is the measure of a user having had an incident of unauthorised access to his/her ICT tool, as well an overall experience such as difficulties managing accounts passwords or accessing ICT tools using passwords.

The results like that of Ugwu *et al*., (2022) and Ugwu *et al.;* (2023) did not establish any significant relationship between educational qualification, and gender to password usage experience in Southeastern Nigeria of respondents under the age of 60. To the best of our knowledge, this is the first research effort towards measuring the experience of end-users to password-based authentication in developing economies alongside relevant factors such as age, educational qualifications, gender, cyber hygiene culture and aligning it with that of end-users' experiences in developed economies.

## 1.1 End-User Experiences With Password-Based Authentication Systems

The use of computers and the Internet on a daily basis has changed the way people conduct their lives and businesses (Butler and Butler, 2015; Shaikh and Karjaluoto, 2015). People can now work using their computer systems from anywhere they can access the Internet (Brynjolfsson *et al*., 2020). These systems are used for a variety of activities such as entertainment (Butler and Butler, 2015), shopping (Vasić *et al*., 2019), banking (Singh and Srivastava, 2020), healthcare services (Liu *et al*., 2019), and maintaining communication on social media (Huerta-Álvarez *et al*., 2020). Both the young and older adults acknowledged the benefits and importance of using technology to accomplish everyday tasks (Morrison *et al*., 2021). One of the significant benefits is convenience, as it allows people to stay and work independently, and many are keen to continue using technology well into older age. However, there have been concerns on the escalation of cybercrime incidents, with consumers prioritising having safe Internet experience, as 55% cite security as the essential aspect of their online experience (BusinessWire, 2021). Similarly, another consumer experience reported by Beyond Identity (2021) reveals that 46% of United States (US) consumers failed to complete transactions due to authentication failure. Lance (2021) shows the circumstances people forget their passwords; (67% of the respondents said it happens when they are trying to complete an online banking transaction, 56% said it happens when trying to get travel information, 55% reported it happens when they are attempting to buy something, and 43% said it happens when they try to access a document).

For decades, user identification and authentication, especially password systems, have been regarded as the foundation of frontline defense against intruders' within a computer security environment (Conklin *et al*., 2004: 1, Güven, Boyaci, & Aydin, 2022). End user identification and authentication in ICT tool through unique username and password have become acceptable, understandable, and even expected to ensure a secure environment (Butler and Butler, 2015). Yang (2019) and Güven, Boyaci, & Aydin, (2022) affirm that user identification and authentication are essential to ensure computer security and control access while maintaining the users' integrity and confidentiality. Although other user authentication systems, such as one-time Personal Identification Numbers (PINs) (using device ownership) and Biometrics (using physical characteristics), Multi-Factor and Two-Factor authentications are evolving, but password-based authentication systems remain one of the most cost-effective and efficient methods to use (Butler and Butler, 2015; Yang, 2019; Tam *et al*., 2010). However, while the authentication of users is critical to controlling access, the authentication process remains problematic (Chiasson and Biddle, 2007; Butler and Butler, 2015).

In addition, Butler and Butler (2015) reveal that because users now have unlimited online options, they are impatient with the time-consuming and inconvenient login experiences. These experiences include being prompted to reset a password or create an account with a long-form and managing passwords (measures related to the safekeeping of passwords).

## 1.2 Password Experiences of Social Media Users

In recent years, social media have gained substantial popularity as it allows individuals to connect, share and network all over the world (Huerta-Álvarez *et al*., 2020). The emergence of Twitter in 2006, YouTube in 2005, Facebook in 2004, and LinkedIn in 2003 has been unprecedented, thus making them a global phenomenon (Rozaimee *et al*., 2013; Tess, 2013). Facebook appears to be the largest and the most populous social networking site for all ages, with over one billion users worldwide (Zeevi, 2013). According to Dunphy *et al.* (2015: 142), social media use by Americans shows that approximately 18% of all online adults are Twitter users, compared to the 71% being users of Facebook, the largest online social platform. While Facebook is popular across a range of demographic groups, 31% of Twitter users are drawn from the age range 18-29 (19% of Twitter users are in the age range 30-49 years; 9% for 50-64 years; 5% for 65+ years) with a particular presence of urban dwellers, African-Americans, and Hispanics. Moreover, Twitter users are split equally in gender, and 46% of all Twitter users tend to visit the site daily (29% multiple times per day).

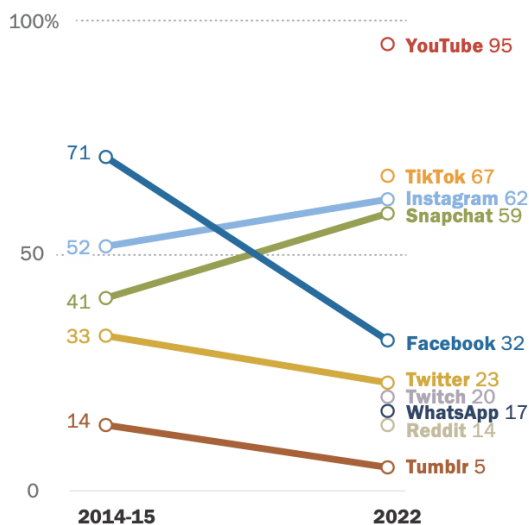

*Figure 1*: *American Teens Social Media and Technology Survey [Vogel et al., 2022]*

Recent survey by Vogels, E, Gelles-Watnick, R and Massarat, N. (2022) of Pew Research Center on America teens between the ages of 13 to 17 shows that TikTok is the most popular social media network, while the use of Facebook by this age group has fallen sharply (Figure 1).

As social networking sites are increasingly becoming an integral part of our everyday lives, it is essential to understand individual experiences as it relates to password-based authentication while using social media platforms to network. For example, Majid and Kouser (2019) report that many users forget to log out of their social media accounts. Likewise, Stobert and Biddle (2014) reveal that several users were more careful about logging out on their computers. This implies that when another person starts using the same mobile/laptop, s/he gets access to the account information and can change the password or even post items and communicate with your friends as if they are from your side. Another user experience shared by Zintle (2020) reveals that, in South Africa, a parliament meeting hosted via the online streaming service Zoom was hacked, and pornographic images were broadcasted. This can only happen when there is a security breach which can also occur if the user's password falls into the wrong hands (Isobe and Ito, 2021).

## 1.3 Password Experiences of Online Shoppers

The online shopping experience continues to evolve as consumers increasingly rely on their social connections (Bilgihan *et al*., 2016). Online customers' experience, thus, includes every point of contact (apps, social media, website) that customer chooses to use to interact and transact with the firm (Bilgihan *et al*., 2016). Most often, online customers generally utilise password authentication when transacting; hence, several scholars have reported consumers' experiences regarding password authentication (Meter and Bauman, 2015; Fagan *et al*., 2017; Beyond Identity, 2021). For example, Beyond Identity (2021) reveals that 46% of US consumers failed to complete transactions due to authentication failure. Also, 18.75% of returning users abandon the cart after forgetting their password and having issues with password reset emails. Likewise, 36% of the consumers said they try to guess a forgotten password twice before resetting it, while 28% said they guess a forgotten password once and 22% three times (Beyond Identity, 2021).

## 1.4 Password Experience of Electronic Healthcare Clients

Some end-users have embraced virtual or electronic healthcare (eHealth) services as a preferable alternative to traditional doctor appointments (Han *et al.*, 2006). They connect with their doctors via phone calls and video chat platforms such as Zoom, Skype, and WhatsApp for consultations without necessarily having physical contact with their doctors (Chatterjee, 2022). However, the underlying issue is security requirements as it relates to eHealth user authentication and authorisation. End-user authentication is very important to ensure confidentiality, which is very essential in the healthcare domain (Chatterjee, 2022: 285). Constantinides *et al.,* (2020) reveal that, in an emergency, the end-users do not have time to reset their passwords when they forget. Similarly, a health worker does not have time to type several codes when accessing the electronic medical records (EMR). Likewise, the end-users are impatient to reset passwords when they forget their passwords (Constantinides *et al.*, 2020).

## 1.5 Password Experience of Electronic Bank Customers

According to Fagan *et al*., (2017), usernames and passwords are still very much ingrained in the fabric of online banking and commerce as the primary means of initial authentication despite being high-risk. Krol *et al.,* (2015) reveal that users expressed much frustration about providing additional information when using online banking; in particular, they did not like to use one-time passwords (OTP) with devices (hardware token). Likewise, Singh *et al.,* (2007) describe users' experiences of how an entire village would delegate bank credentials to a single person who would conduct the online transaction on everybody's behalf. In addition, Dunphy *et al*., (2014: 2) report on how older adults delegate the withdrawal of their monies from an automated teller machine to their helpers thereby disclosing their personal identification numbers (PINs) to a third party.

## 1.6 Research Model

Several Information System (IS) models have been used to study users' experiences in adopting new technologies. A typical example is the Unified Theory of Technology of Acceptance and Use of Technology (UTAUT) (Venkatesh *et al.*, 2003). It holds that four constructs, performance expectancy, effort expectancy, social influence, and facilitating conditions, directly influence user behaviour. It has moderating factors of gender, age, experience, and voluntariness of use. However, some constructs of the UTAUT, such as social influence and performance expectancy, are not significant because this study does not intend to measure performance expectancy but to explore the experiences of online end-users in using a password-based authentication system. Nevertheless, this study adopts the moderating factors of age, educational level, and gender as external variables in the conceptual framework. Thus, this study aims to measure online end-users' experiences as it relates to password-based authentication system usage. The research article attempts to answer the following hypothesis:

**H1**: Consumers' experiences have a positive influence on the usage of password-based authentication systems.

**H0**: Consumers' experiences have a negative influence on the usage of password-based authentication systems.

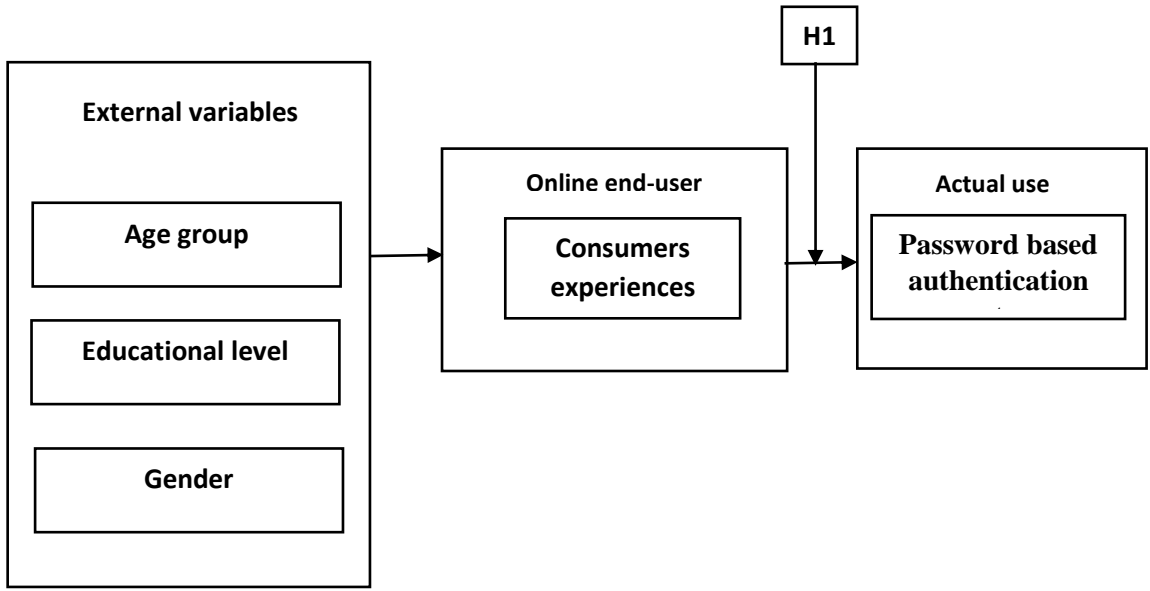

*Figure 2: Variables contributing to online consumers' experiences while using a Password-based authentication system [Source: Primary]*

### 1.6.1 External Variables

The external variables contribute to consumers' experiences while transacting online using a password-based authentication system, as shown in Figure 2. These variables include age group, educational level, and gender. Age group can be described as one of the criteria used to include or exclude certain audiences (Natarajan *et al*., 2018). In this study, the age group helps determine the experiences of online end-users when transacting via a password-based authentication system. While educational level helps to measure the end-user experiences when transacting via a password-based authentication system. Likewise, gender helps to understand the influence of the end-users' gender differences as it relates to their experiences while transacting through a password-based authentication system. For this study, the chosen external variables will represent unique characteristics of the sample that could have impact on the online end-users using a password-based authentication system (Venkatesh *et al*., 2003).

### 1.6.2 Online End-User Behaviour

The external variables influence the end-user's behaviour construct. From the research, the experiences that the consumer displays due to external variables could give a user a more positive/negative attitude towards using the password-based authentication system *(*Venkatesh *et al.,* 2012). According to Brock and Khan (2017) behavioural intention to use is an essential step towards the actual usage of any new system. In this study, online end-users' behaviour will be used to evaluate online consumers' experiences while using a password-based authentication system.

### 1.6.3 Actual Use

Actual use of a system/technology is determined by users behaviour (Venkatesh *et al*., 2012). In this study, actual use indicates the end-user's usage of a password-based authentication system while transacting online. The end-user experiences will determine the predictions of the usage. According to Butler and Butler (2015), many computer security password breaches result from poor user security behaviour. The password creation and management practices that online consumers apply have a direct effect on the level of computer security and are often targeted in attacks.

## 2.0 Research Methodology

This study was conducted using some Universities within the southeastern region of the country with ethical approval obtained from one of the concerned authorities. The study used online-administered survey approach and the questionnaire was designed in Google form and distributed to our respondents through WhatsApp and email addresses. Data from respondents were coded into numerical data in a spreadsheet. Processing and analysis of captured data was carried out using SPSS (Statistical Package for Social Sciences) version 20.0 and reported by means of descriptive statistics, frequency distribution, and Chi-Square tests. In all, about 291 responses were used after data cleaning in accordance to the required sample size of the respondent population (Krejcie & Morgan (1970)).

### 2.1  Instruments and Methods

For the researchers to verify the hypotheses formulated, quantitative research methodology was adopted in the study which according to Adegbuyi *et al*., (2015) is the most suitable for any exploratory investigation for better understating of a particular problem under scrutiny. The variables needed to measure the main constructs that are part of the conceptual framework were identified through extensive review of literature. A cross-sectional research design in conjunction with a well-structured questionnaire with closed-ended questions as used by Bowen *et al*., (2010) was adopted to gather information from the respondents. The questionnaire is comprised of two sections. Section A has questions that focused on obtaining the demographics of the respondents such as age, gender, educational qualification, marital status, and employment information while section B examined the online end-user experiences with the Password-based authentication systems. The questions in section B were designed in such a way that data can be collected based on the respondents' experiences with Password-based authentication systems.

### 2.2 Sample Size and Sampling Technique

A population of 1200 possible respondents were drawn from Universities in the Southeastern Nigeria for the purpose of this study with an expected sample size of 291 as stipulated in Krejcie & Morgan (1970). Since the possibility of collecting data from the entire population is not possible according to Sekaran & Bougies, (2010), using sample relative to the entire population was considered as being able to produce reliable and better result and remove unnecessary stress. A systematic sampling approach was adopted to provide for a heterogenous population characteristics (Leedy & Ormrod, 2005) and to make generalisation to the population (Bryman & Bello, 2003). Systematic sampling technique was considered suitable by previous studies such as (Albuelu & Ogbouma, 2013; Asgharnezhad, Akbarlou & Karlcaj, 20 13; Aliyu, 2013). At the end of the survey, a sample size of 291 was used for the study after data cleaning, which represented the actual number as suggested by Krejcie & Morgan (1970).

## 2.3 Data Collection Method

The study used online-administered survey approach. The questionnaire was designed in Google form and distributed to our respondents through WhatsApp and email addresses. The primary targets of respondents were students and employees of the Universities since they were perceived likely to be the most informed users of online Password-based authentication system. Prior to the distribution of the questionnaire, an ethical approval was obtained for the study from the approved authority. The responses to the questionnaire were collected through the spreadsheet designed for this purpose. A total of 299 responses were received out of the target population of 1200 and were subjected to data cleaning and normalisation. The extracted and normalised data were subsequently subjected to analysis.

## 3.0 Analysis, Results and Discussion
## 3.1 Data Analysis

Data from respondents were coded into numerical data in a spreadsheet. Processing and analysis of captured data was carried out using SPSS (Statistical Package for Social Sciences) version 20.0 and reported by means of descriptive statistics, frequency distribution, and Chi-Square tests.

## 3.2 Results

The demographic data is presented in Table 1 below. The table shows that 157 male respondents (54%) and 134 female respondents (46%) participated in the survey. Ages of the participants were divided into five categories. Participants that are 20 years and below has a distribution of 86 out of 291 participants, which represents 29.6% of all participants. The second category captured participants that are between 21-30. This group has a distribution of 148(50.9%) of 291 respondents. The third category captured respondents between 31-40 years and had a distribution of 28 representing 9.6% of all participants. In category 4, respondents between 41-50 were captured with a distribution of 21, representing 7.2% of entire respondents. The last category captured respondents within the ages of 51-60. A distribution of 8 was observed, representing 2.7% of the entire participants. The third demographic data collected is the highest education qualification obtained by the respondents. The first category captured are respondents with secondary school certificate. 187 respondents out of the 291, representing 64.3% of participants belong to this category. The second category had OND (Ordinary National Diploma) certificate owners covered. This category had a distribution of 16 respondents, which is 5.5% of the entire 291 respondents. Other categories are HND (Higher National Diploma and B.Sc. (Bachelor of Science) with distributions of 42 (14.4%) and those with Ph.D. (Doctor of Philosophy) had a distribution of 46 (15.8%) respectively. Demographic data on the employment status of respondents showed that students/unemployed respondents had a distribution of 243 out of 291 participants, giving 83.5% of all participants. Contract/temporary staff had a distribution of 3 (1.0%), while permanent staff and undefined categories had a distribution of 39 (13.4%) and 6 (2.1%) respectively.

**Table 1:** Demographic Information

| Demographic Data | Frequency | Percent | Cumulative Percent |
|---|---|---|---|
| **Gender** | | | |
| Male | 157 | 54.0 | 54.0 |
| Female | 134 | 46.0 | 100.0 |
| **Age** | | | |
| <=20 | 86 | 29.6 | 29.6 |
| 21-30 | 148 | 50.9 | 80.4 |
| 31-40 | 28 | 9.6 | 90.0 |
| 41-50 | 21 | 7.2 | 97.3 |
| 51-60 | 8 | 2.7 | 100.0 |
| **Educational Qualification** | | | |
| School Certificate | 187 | 64.3 | 64.3 |
| OND (ND) | 16 | 5.5 | 69.8 |
| HND/B.Sc. | 42 | 14.4 | 84.2 |
| Ph.D. | 46 | 15.8 | 100.0 |
| **Employment Status** | | | |
| Students/Not Employed | 243 | 83.5 | 83.5 |
| Contract/Temporary Staff | 3 | 1.0 | 84.5 |
| Permanent Staff | 39 | 13.4 | 97.9 |
| Other | 6 | 2.1 | 100.0 |

Chi-Square tests were used to gain insight into the association level that exists between the variables.

**Assumptions**: for a 2 X 2 table, not more than 10% of cells have expected counts of less than 5, while for tables greater than 2 X 2, not more than 20% of cells should have expected counts of less than 5. If the assumption is violated, the readings are taken from the Likelihood Ratio row instead of the Pearson Chi-Square row.

**Research Question 1**
What are the end-user experiences in password-based authentication?

To answer this question, descriptive analysis of respondents' self-reports on variables in this domain was carried out. Tables 2 and 3 below show the results of the analysis.

**Table 2**: Frequency Distribution of Respondents (Experienced unauthorized access to account?)

| | | Frequency | Percent | Cumulative Percent |
|---|---|---|---|---|
| Valid | Yes | 93 | 32.0 | 32.0 |
| | Not At All | 153 | 52.6 | 84.5 |
| | Don't Know | 25 | 8.6 | 93.1 |
| | Can't Remember | 20 | 6.9 | 100.0 |
| | **Total** | **291** | **100.0** | |

From Table 2 above, 93 respondents, giving a distribution of 32% of total respondents, have experienced unauthorised access to their accounts, while a higher percentage of the respondents 153 (52.6%) have not. Also, those who did not notice if such compromise occurred and those who cannot remember its occurrence are 25 (8.6%) and 20(6.9%) respectively.

To further ascertain the respondents' password-based authentication experiences, the respondents were presented with the question: **How has your experience been using Password authentication?** Table 3 below gives the self-report frequency:

**Table 3**: Frequency Distribution of Respondents End-User Experience (Rating)

|  |  | Frequency | Percent | Cumulative Percent |
|---|---|---|---|---|
| Valid | Excellent | 126 | 43.3 | 43.3 |
|  | Good | 139 | 47.8 | 91.1 |
|  | Average | 23 | 7.9 | 99.0 |
|  | Poor | 3 | 1.0 | 100.0 |
|  | **Total** | **291** | **100.0** |  |

Only 3 (1%) of the 291 respondents reported having a poor experience with Password authentication while 126 (43.3%) reported excellent Password authentication experience, 139 (47.8%) and 23 (7.9%) reported good and average experiences respectively. This agrees with the report from Table 2, where majority of the respondents reported not having had their accounts compromised, and either not being aware of such compromise or cannot remember its occurrence.

**Research Question 2**
Does age have impact on end-user experience in Password-based authentication?

To answer this question, a hypothesis was raised and tested.
Null Hypothesis(**H0**): Age of Internet users have no impact on their Password-based authentication experiences.

**Hypothesis Testing**
Cross tabulation of the independent variable (age) and dependent variable (Password-based authentication experience) showed observed and expected counts recorded on each category's option item and assumption check following Chi-Square tests is used to select that appropriate values required to ascertain the association strength between the variables and if the independent impacts on the dependent.

**Table 4**: Age * End-user experience in Password-based authentication (Has someone ever logged into your account without your permission?)

| Age |  | Has someone logged into your account without your permission? | | | | Total |
|---|---|---|---|---|---|---|
|  |  | Yes | Not At All | Don't Know | Can't Remember |  |
| <=20 | Count | 31 | 40 | 9 | 6 | 86 |
|  | Expected Count | 27.5 | 45.2 | 7.4 | 5.9 | 86.0 |
|  | % Within Age | 36.0% | 46.5% | 10.5% | 7.0% | 100.0% |
| 21-30 | Count | 47 | 82 | 11 | 8 | 148 |
|  | Expected Count | 47.3 | 77.8 | 12.7 | 10.2 | 148.0 |
|  | % Within Age | 31.8% | 55.4% | 7.4% | 5.4% | 100.0% |
| 31-40 | Count | 8 | 16 | 1 | 3 | 28 |
|  | Expected Count | 8.9 | 14.7 | 2.4 | 1.9 | 28.0 |
|  | % Within Age | 28.6% | 57.1% | 3.6% | 10.7% | 100.0% |
| 41-50 | Count | 4 | 12 | 3 | 2 | 21 |
| | Expected Count | 6.7 | 11.0 | 1.8 | 1.4 | 21.0 |
| | % Within Age | 19.0% | 57.1% | 14.3% | 9.5% | 100.0% |
| 51-60 | Count | 3 | 3 | 1 | 1 | 8 |
| | Expected Count | 2.6 | 4.2 | 0.7 | 0.5 | 8.0 |
| | % Within Age | 37.5% | 37.5% | 12.5% | 12.5% | 100.0% |
| Total | Count | 93 | 153 | 25 | 20 | 291 |
| | Expected Count | 93.0 | 153.0 | 25.0 | 20.0 | 291.0 |
| | % Within Age | 32.0% | 52.0% | 8.6% | 6.9% | 100.0% |

The cross-tabulation result shows the disparity between the observed counts and the expected counts. The expected counts are counts that would be observed if no association exists between the dependent and independent variables. The greater the difference between these counts, the higher the impact independent variable has on the dependent variable.

In age category <=20, count of 31(36.0%) was observed for respondents who have experienced unauthorized access to their accounts whereas the expected count was 27.5(approximately 28). 40(46.5%) respondents were observed as having not experienced unauthorized access where the expected count was 45.2(approximately 45). Option items "Don't know" and "Can't remember" had observed counts of 9(10.5%) and 6(7.0%) respectively, against expected counts of 7.4 (approximately7) and 5.9 (approximately 6).

The second age category captured were respondents between ages 21-30. In this category, observed counts of 47(31.8%) were observed for respondents who have experienced unauthorised access to their accounts against expected count of 47.3(approximately 47). 82 (55.4%) counts were observed instead of an expected count of 77.8 (approximately 78) for respondents who have not experienced unauthorised access to their accounts, while 11(7.4%) and 8 (5.4%) counts were observed instead of expected 12.7(approximately 13) and 10.2(approximately 10) respectively, for respondents who either have no knowledge of unauthorised access to their accounts or cannot remember its occurrence.

The third age category, 31-40 had observed counts of 8(28.6%) for unauthorised access experience, with expected count of 8.9(approximately 9), 16(57.1%) observed counts for no unauthorised access experience, with an expected count of 14.7(approximately 15), 1(3.6%) and 3(10.7%) observed counts and of 2.4(approximately 1) and 1.9 (approximately 2) expected counts respectively were recorded for respondents who do not know of any attempt to access their accounts without permission or cannot remember its occurrence.

Age category four captured respondents between the ages of 41-50. Here, 4(19.0%) respondents have had their accounts accessed without their permission while an expected count of 6.7 (approximately 7) would be recorded when no association exists between age and end-user experience. 12(57.1%) observed counts and expected 11 counts for respondents who have not had their accounts accessed by unauthorised persons. Observed counts for respondents who either do not know of unauthorised access to their accounts or cannot remember its occurrence are 3(14.3%) and 2 (9.5%), with 1.8 (approximately 2) and 1.4(approximately 1) as expected counts in these option items respectively.

The last age category captured respondents between ages 51 to 60. A distribution of 3(37.5%) observed to 2.6(approximately 3) expected counts for Yes to unauthorised account access, 3 (37.5%) observed to 4.2 (approximately 4) expected counts for no experience of unauthorised access, 1 (12.5%) observed to 0.7(approximately 1) expected count for respondents who do not know of the occurrence of unauthorised access to their accounts and 1(12.5%) observed count to 0.5 (approximately 1) expected count for respondents who cannot remember having their accounts accessed by unauthorised persons.

**Assumption:** The table is greater than 2 x 2, and the assumption is that not more than 20% of expected count is greater less than 5. This assumption applies to Table 4, which is a 4 x 5 table (4 categories by 5 categories).

For each of the variables, a chi-square test of independence was used to test for a significant relationship between "Age" and "End-user experience in Password-authentication (Has someone ever logged into your account without your permission?)". Chi-Square test result is displayed on Table 5 below.

**Table 5:** Chi-Square Tests

|  | Value | Df | Asymp. Sig(2-sided) | Exact Sig. (2 sided) | Exact Sig.(1-sided) |
|---|---|---|---|---|---|
| Pearson Chi-Square | 7.080 [a] | 12 | 0.852 | b |  |
| Likelihood Ratio | 7.227 | 12 | 0.842 |  |  |
| Fisher's Exact Test | 8.751 |  |  | 0.683 | b |
| Linear-by-Linear Association | 1.369 | 1 | 0.242 |  |  |
| N of Valid Cases | 291 |  |  |  |  |

a. **8 cells (40.0%) have expected count less than 5. The minimum expected count is 0.55**
b. Not computed

From Table 5, the result shows that 40% (8 cells) have expected counts of less than 5, hence violating the assumption that not more than 20% of cells should have expected counts of less than 5. When the assumption is violated, Pearson Chi-Square row values are ignored, and the Likelihood Ratio row values are used to determine the statistical significance and impacts level of the independent variable on the dependent variable.

With regards to the difference between the observed counts and expected counts, it is obvious that an association exists between the independent variable "Age" and the dependent variable "End-user Password-authentication experience (Has someone logged into your account without your permission?)". The likelihood Ratio reading gives:

**Statistical Value**: 7.227; **Degree of Freedom:** 12; **P-Value**: 0.842, represented on the table as Asymptotic 2-sided.

The *p*-value reflects the degree of data compatibility with the null hypothesis. Also, 0.05 (5%, *i.e.*, 1 of 20) has been conventionally accepted as the threshold to differentiate significant from non-significant results (Di Leo and Sardanelli, 2020).

In this result (Table 5), the reported p-value is 0.842, which is much larger than the significance level of 0.05. This means the result of tests are not statistically significant at 0.05 level. In other words, we cannot reject the null hypothesis (**H0**). We therefore accept the Null Hypothesis (**H0**), which states that the age of respondents has no impact on their end-user Password authentication experience (experienced unauthorised access to account).

It is important to note that when conducting a hypothesis test, the null hypothesis is typically rejected if the p-value is less than the significance level. The significance level is the probability of rejecting the null hypothesis when it is true. The most used significance level is 0.05, which means that there is 5% chance of rejecting the null hypothesis when it is true.

Again, another test was carried out using the hypothesis with another variable that captures user experience, scaled by option items that allow Internet users to rate their experience with Password authentication: the result is displayed on table 6 below:

**Table 6**: Chi-Square tests for association between Age and End-user experience (How has your experience been using Password authentication?)

| Age | | How has your experience been using Password authentication? | | | | Total |
|---|---|---|---|---|---|---|
| | | Excellent | Good | Average | Poor | |
| <=20 | Count | 32 | 47 | 7 | 0 | 86 |
| | Expected Count | 37.2 | 41.1 | 6.8 | 0.9 | 86.0 |
| | % Within Age | 37.2% | 54.7% | 8.1% | 0.0% | 100.0% |
| 21-30 | Count | 65 | 69 | 13 | 1 | 148 |
| | Expected Count | 64.1 | 70.7 | 11.7 | 1.5 | 148.0 |
| | % Within Age | 43.9% | 46.6% | 8.8% | 0.7% | 100.0% |
| 31-40 | Count | 18 | 8 | 2 | 0 | 28 |
| | Expected Count | 12.1 | 13.4 | 2.2 | 0.3 | 28.0 |
| | % Within Age | 64.3% | 28.6% | 7.1% | 0.0% | 100.0% |
| 41-50 | Count | 8 | 11 | 1 | 1 | 21 |
| | Expected Count | 9.1 | 10.0 | 1.7 | 0.2 | 21.0 |
| | % Within Age | 38.1% | 52.4% | 4.8% | 4.8% | 100.0% |
| 51-60 | Count | 3 | 4 | 0 | 1 | 8 |
| | Expected Count | 3.5 | 3.8 | 0.6 | 0.1 | 8.0 |
| | % Within Age | 37.5% | 50.0% | 0.0% | 12.5% | 100.0% |
| Total | Count | 126 | 139 | 23 | 3 | 291 |
| | Expected Count | 126.0 | 139.0 | 23.0 | 3.0 | 291.0 |
| | % Within Age | 43.3% | 47.8% | 7.9% | 1.0% | 100.0% |

Table 6 above shows the observed and expected counts resulting from the cross tabulation of Age of respondents (independent variable) and Password authentication experiences (dependent variable). The dependent variable is however captured using a question that allows respondents select their Password authentication experiences from given option items.

Cross tabulation of the two participating variables give insight into the existence of association or otherwise between age of respondents and their Password authentication experiences.

The distribution shows that for respondents in age category <=20, count of 32(37.2%) was observed for respondents who have excellent Password authentication experience whereas the expected count was 37.2(approximately 37). 47(54.7%) observed counts were recorded for respondents who reported having good Password authentication experiences where the expected count was 41.1(approximately 41). The other options (average and poor Password authentication experiences had observed counts of 7 (8.1%) and 0(0%) against expected counts of 6.8 (approximately 7) and 0.9 (approximately 1) respectively.

The second age category captured respondents between ages 21-30. In this category, observed count of 65(43.9%) was recorded for respondents who have excellent Password authentication experiences against expected count of 64.1(approximately 64). 69(46.6%) observed counts were recorded for respondents with good Password authentication experiences, with the expected count being 70.7(approximately 71). 13(8.8%) and 1(0.7%) count were observed for respondents who have average and poor Password authentication experiences respectively. The expected counts for the average and poor experiences are 11.7(approximately12) and 1.5 (approximately 2) respectively.

The third age category, 31-40 had observed counts of 18(64.3%) for respondents whose Password authentication experiences are excellent, with expected count of 12.1(approximately 12). 8(28.6%) respondents reported having good authentication experiences, with an expected count of 13.4, 2(7.1%) and 0(0.0%) observed counts with expected counts of 2.2(approximately 2) and 0.3 (approximately 0) for respondents who have average and poor Password authentication experiences respectively. Expected counts were recorded for respondents who have average and poor experiences respectively.

Age category four captured respondents between the ages of 41-50. Here, 8(38.1%) observed counts were obtained for respondents who have excellent authentication experiences with an expected count of 9.1(approximately 9). Observed counts for respondents who have average and poor experiences are 1(4.8%) and 1(4.8%) respectively, with expected counts of 1.7(approximately 2) and 2 respectively.

The last age category, 51-60 has observed count of 3(37.5%) for respondents who consider their Password authentication experience excellent, with expected counts of 3.5(approximately 4). 4(50.0%) observed counts were recorded in the good experience domain, with 3.8 (approximately 4) expected counts. No observed count was recorded for average Password authentication experience. However, 0.6(approximately 1) expected count correspond to this option item. 1(12.5%) observed count to 0.1 expected for respondents whose Password authentication experiences are poor.

Following the disparity between observed and expected counts, the assumption was evaluated. The assumption for using Pearson Chi-Square reading is that not more than 20% of expected counts are less than 5. Table 7 below shows the output of the test.

**Table 7:** Chi-Square Tests

|  | Value | Df | Asymp. Sig.(2-sided) | Exact Sig. (2 sided) | Exact Sig.(1-sided) | Point Probability |
|---|---|---|---|---|---|---|
| Pearson Chi-Square | 22.413 [a] | 12 | 0.033 | [b] |  |  |
| Likelihood Ratio | 15.889 | 12 | 0.196 | 0.188 |  |  |
| Fisher's ExactTest | 15.889 |  |  | 0.145 |  |  |
| Linear-by-Linear Association | 0.15 [c] | 1 | 0.903 | 0.927 | 0.473 | 0.036 |
| N of Valid Cases | 291 |  |  |  |  |  |

a. 10 cells (50.0%) have expected count less than 5. The minimum expected count is 0.08
b. Not computed
c. Standardized statistics is -0.122

Sequel to the difference between the observed counts and expected counts, it is obvious that an association exists between the independent variable "Age" and the dependent variable "End-user Password-authentication experience, scaled according to strength of users' perceived feeling of satisfaction (excellent, good, average, or poor). The table shows that 50% of cells have counts less than 5 hence the assumption for using the Pearson Chi-Square reading row has been violated. We therefore take the reading of the Likelihood Ratio row in ascertaining the statistical relevance of the existing association.

**Statistical Value:** 15.889**; Degree of Freedom:** 12**; P-Value:** 0.196**.**

The result on Table 7 represented by Asymptotic 2-sided significance shows the p-value is 0.196. However, the acceptable statistical significance level is 0.05, which means that the p-value 0.196 is greater than the significance level. This implies that the result of the tests are not statistically significant at 0.05 level and we cannot reject the null hypothesis. The p-value of 0.196 is above the 0.05 value acceptable for statistical significance, hence we accept the **H0** which states that age has no impact on end-user Password authentication experience. We therefore conclude that age of respondents has no impact on their Password authentication experience.

### Research Question 3
Does gender have impact on End-user experience in Password-based authentication?

To answer this research question, a hypothesis was raised and tested:
Null Hypothesis(**H0**): Gender of Internet users has no impact on their Password authentication experiences.

In this section we discuss the result from analysis of the cross tabulation of independent variable "Gender" and Dependent variable (End-User Password Authentication Experience), captured by questionnaire item: has someone ever logged into your account without permission? The association is tested using Chi-Square cross tabulation of the two variables which reveal the counts observed on each option item and compared the observed count to an expected count that would be obtained in the absence of an association between the independent and dependent variables. Table 8 below give the crosstab output:

**Table 8:** Gender * User Experience (Has someone logged into your account without your permission?)

| GENDER | | Has someone logged into your account without your permission? | | | | Total |
|---|---|---|---|---|---|---|
| | | Yes | Not At All | Don't Know | Can't Remember | |
| MALE | Count | 46 | 88 | 9 | 14 | 157 |
| | Expected Count | 50.2 | 82.5 | 13.5 | 10.8 | 157.0 |
| | % Within Age | 29.3% | 56.1% | 5.7% | 8.9% | 100.0% |
| FEMALE | Count | 47 | 65 | 16 | 6 | 134 |
| | Expected Count | 42.8 | 70.5 | 11.5 | 9.2 | 134.0 |
| | % Within Age | 35.1% | 48.5% | 11.9% | 4.5% | 100.0% |
| Total | Count | 93 | 153 | 25 | 20 | 291 |
| | Expected Count | 93.0 | 153.0 | 25.0 | 20.0 | 291.0 |
| | % Within Age | 32.0% | 52.0% | 8.6% | 6.9% | 100.0% |

From the table above, 46 male respondents (29.3%) have experienced unauthorized access to their accounts, where expected count of 50.2 (approximately 50) would be recorded where no association exists between Gender and End-user experience (Has someone logged into your account without your permission). 88 male respondents (56.1%) have not had their accounts accessed by unauthorized persons. However, this count would be 83 assuming no association existed between gender and end-user experience (Has someone logged into your account without your permission). Male respondents who selected option item "Don't know" have observed counts of 9(5.7%), with expected count of 3.5 (approximately 4).

The fourth option item "Can't remember" has an observed count of 14 (8.9%) with expected counts of 10.8 (approximately 11). 47 (35.1%) of the female respondents have had their accounts accessed by unauthorized persons, with expected count of 42.8 (approximately 43); 65 (48.5%) observed counts for respondents that have not experienced unauthorized access to their account with an expected count of 70.5 (approximately 71). The third option item and fourth options items in the female gender category (Don't know and can't remember) had observed counts of 16 (11.9%) and 6 (4.5%) respectively, with expected counts of 11.5(approximately 12) and 9.2(approximately 9) assuming no association exists between the independent variable "Gender" and the dependent variable "End-user Password authentication experience (Has someone logged into your account without your permission).

From the disparities between observed and expected counts, an association exists between gender and end-user Password authentication experience, however, to test the strength of the association, Chi-Square test result would be examined. The assumption for using the Pearson Chi-Square is that not more than 20% of expected count is less than 5. Table 9 below shows the Chi-Square result:

**Table 9**: Chi-Square Tests

|  | Value | Df | Asymp. Sig(2-sided) | Exact Sig. (2 sided) | Exact Sig.(1-sided) | Point Probability |
|---|---|---|---|---|---|---|
| Pearson Chi-Square | 6.853 [a] | 3 | 0.077 | 0.075 |  |  |
| Likelihood Ratio | 6.939 | 3 | 0.74 | 0.077 |  |  |
| Fisher's ExactTest | 6.746 |  |  | 0.078 |  |  |
| Linear-by-Linear Association | 0.766 [b] | 1 | 0.381 | 0.391 | 0.211 | 0.039 |
| N of Valid Cases | 291 |  |  |  |  |  |

a. 0 cells (0.0%) have expected count less than 5. The minimum expected count is 9.21
b. The Standardized statistic is: -0.875

From the table above, 0 (0.0%) cells have expected counts less than 5, with the minimum expected count being 9.21. this implies that the assumption for reading the result from the Pearson Chi-Square row is met hence the following readings:

**Statistical Value:** 6.853; **Degree of Freedom:** 3; **P-Value:** 0.077

Accepted statistical significance level is 0.05, hence the p-value from this test, represented on the table by Asymptotic Significance (2-sided), is 0.077. This result is more than the accepted level. In other words, the null hypothesis cannot be rejected as there is no enough evidence to suggest that the alternative hypothesis (H1) is true. We therefore accept the Null Hypothesis (**H0**): Gender of Internet users do not have impact on their end-user experience in Password-based authentication and reject the alternative hypothesis (H1): Gender of Internet users have impact on their end-user experience in Password-based authentication.

Further testing of the impact gender has on end-user Password authentication experience was carried out using respondents self-report on their authentication experiences, scaled according to their perceived rating of the experience. Table 10 below displays the crosstab result of respondents' gender and their self-rated Password authentication experience.

**Table 10:** Gender * User Experience (How has your experience been using Password authentication?)

| Gender | | How has your experience been using Password authentication? | | | | Total |
|---|---|---|---|---|---|---|
| | | Excellent | Good | Average | Poor | |
| Male | Count | 70 | 71 | 14 | 2 | 157 |
| | Expected Count | 68.0 | 75.0 | 12.4 | 1.6 | 157.0 |
| | % Within Age | 44.6% | 45.2% | 8.9% | 1.3% | 100.0% |
| Female | Count | 56 | 68 | 9 | 1 | 134 |
| | Expected Count | 58.0 | 64.0 | 10.6 | 1.4 | 134.0 |
| | % Within Age | 41.8% | 50.7% | 6.7% | 0.7% | 100.0% |
| Total | Count | 126 | 139 | 23 | 3 | 291 |
| | Expected Count | 126.0 | 139.0 | 23.0 | 3.0 | 291.0 |
| | % Within Age | 43.3% | 47.8% | 7.9% | 1.0% | 100.0% |

The table above shows the observed counts and their corresponding expected counts, obtained from cross tabulation of the participating variables: Gender and End-user Password authentication experience. These two counts are used to determine if an association exists between the independent variable Gender and the dependent variable: End-user Password authentication experience. Existence of an association implies that one of the variables has impact on the other.

The degree of the impact, however, determines if the association is statistically significant and able to cause a meaningful impact. 70(44.6%) respondents in the male category reported having excellent Password authentication experiences. This count would be 68 if there was no association between gender and end-user Password authentication experience. 71(45.2%) respondents in this category reported having good Password authentication experiences. This count would be 75 assuming no association existed between gender and end-user Password authentication experience. Respondents in the male category, who consider their Password authentication experiences to be average or poor recorded 14(8.9%) and 2(1.3%) counts respectively with 12 expected average counts and 2 expected poor counts assuming no association exists between the independent variable (gender) and the dependent variable (end-user Password authentication experience).

The female category has 56(41.8%) of respondents that reported having excellent Password authentication experiences, with an expected count of 58 respondents assuming no association exists between gender and end-user Password authentication experience. 68(50.7%) respondents in this category reported having good Password authentication experiences. This count would be 64 where no association exists between gender and end-user Password authentication. Average and poor experiences in the female category were recorded by 9(6.7%) and 1(0.7%) respondent respectively, with expected counts for these option items being 11 for average experience and 1 for poor experience.

The disparities between observed and expected counts give credence to an existing association whose strength determine whether the independent variable (gender) has impact on the dependent variable (End-user Password authentication experience). To test this strength, Chi-Square tests is carried out, assumption is tested, and the insight gained from the test used to make appropriate assertions.

**Assumption:** The assumption for reading from the Pearson Chi-Square states that not more than 20% of cell in a table greater than 2 X 2 should have expected counts below 20%. Table 11 below give the result of the Chi-Square test.

**Table 11**: Chi-Square Tests

|  | Value | Df | Asymp. Sig(2-sided) | Exact Sig. (2 sided) | Exact Sig.(1-sided) | Point Probability |
|---|---|---|---|---|---|---|
| Pearson Chi-Square | 1.230 [a] | 3 | 0.746 | 0.740 |  |  |
| Likelihood Ratio | 1.239 | 3 | 0.744 | 0.740 |  |  |
| Fisher's ExactTest | 1.304 |  |  | 0.748 |  |  |
| Linear-by-Linear Association | 0.003 [b] | 1 | 0.953 | 1.000 | 0.512 | 0.070 |
| N of Valid Cases | 291 |  |  |  |  |  |

a. 2 cells (25.0%) have expected count less than 5. The minimum expected count is 1.38
b. Standardized statistic is -0.059

From the table above, 2 (25.0%) cells have expected counts less than 5, with the minimum expected count being 1.38. This implies that the assumption for reading the result from the Pearson Chi-Square row is violated hence we ignore the Pearson Chi-Square readings and use the reading of the Likelihood Ratio row instead. The following are resulting values from the likelihood Ratio:

**Statistical Value**: 1.239; **Degree of Freedom**: 3; **P-Value**: 0.744

As indicated in Table 11, the P-value is 0.744. This p-value is greater than the significance level of 0.05, hence, we cannot reject the null hypothesis (**H0**). We therefore accept the Null Hypothesis (**H0**): Gender of the Internet users do not have impact on their end-user Password authentication experiences and reject the alternative hypothesis: Gender of Internet users has impact on their end-user experience in Password-based authentication.

### Research Question 4
Does Education Level have impact on end-user experience in Password-based authentication?

To answer this research question, we raise and test a hypothesis.
Null Hypothesis (**H0**): Level of education has no impact on end-user experience in Password-based authentication.

To test this hypothesis, we first carried out a cross tabulation of the independent variable: Level of Education and the dependent variable: End-user Password authentication experience. Table 12 below give details of observed counts and corresponding expected counts in the various categories of education level.

In this section, we analyzed the end-user experience using respondents' self-report on whether or not they have had their accounts compromised.

**Table 12**: Education Qualification * User Experience (Has someone logged into your account without your permission?)

| Highest Education Qualification (HEQ) | | Has someone logged into your account without your permission? | | | | Total |
|---|---|---|---|---|---|---|
| | | Yes | Not At All | Don't Know | Can't Remember | |
| School Certificate | Count<br>Expected Count<br>% Within HEQ | 60<br>59.8<br>32.1% | 96<br>98.3<br>51.3% | 19<br>16.1<br>10.2% | 12<br>12.9<br>6.4% | 187<br>187.0<br>100.0% |
| OND | Count<br>Expected Count<br>% Within HEQ | 7<br>5.1<br>43.8% | 8<br>8.4<br>50.0% | 1<br>1.4<br>6.2% | 0<br>1.1<br>0.0% | 16<br>16.0<br>100.0% |
| HND/B.Sc | Count<br>Expected Count<br>% Within HEQ | 13<br>13.4<br>31.0% | 27<br>22.1<br>64.3% | 1<br>3.6<br>2.4% | 1<br>2.9<br>12.4% | 42<br>42.0<br>100.0% |
| Ph.D | Count<br>Expected Count<br>% Within HEQ | 13<br>14.7<br>28.3% | 22<br>24.2<br>47.8% | 4<br>4.0<br>8.7% | 7<br>3.2<br>15.2% | 46<br>46.0<br>100.0% |
| Total | Count<br>Expected Count<br>%Within HEQ | 93<br>93.0<br>32.0% | 153<br>153<br>52.6% | 25<br>25.0<br>8.6% | 20<br>20.0<br>6.9% | 291<br>291.0<br>100.0% |

OND/ND – Ordinary National Diploma, HND = Higher National Diploma, BSC = Bachelor of Science, PhD = Doctor of Philosophy

Distributions from the table above show that 60(59.8%) respondents with School certificate as their highest obtained qualification have had their accounts accessed without authorization. An expected count of 59.8 (approximately 60) would be expected if no association exists between education qualification and end-user Password authentication experience item. 96 (51.3%) respondents in the school certification category have not had their accounts accessed by unauthorized persons. The expected count for this option item assuming no association exists between education level and end-user Password authentication experience would be 98 counts. School certificate holders who either do not know if their accounts have been accessed without authorization or cannot remember its occurrence have 19(10.2%) and 12(6.4%) counts respectively, with 16 expected counts for respondents who do not know and 13 expected counts for those who cannot remember.

The second category captured respondents with Ordinary National Diploma as their highest obtained qualification. The distribution in this category show that 7(43.8%) of the respondents have had their accessed without authorization and that this count would be 5 if no association exists between education level and end-user Password authentication experience as reflected by unauthorized account access.

8(50.0%) respondents in this category have not had their accounts accessed without authorization. A count of 8.4(approximately 8) is expected where no association exists between education qualification and end-user experience item. 1(6.2%) respondent does not know if his/her account has been accessed without authorization. The expected count of 1.4 (approximately 1) would also be obtained where no association exists between education level and end-user Password authentication experience. The last option item in this category has no observed count, but 1 count is expected in event of no association between education level and end-user Password authentication item.

In the third category, Higher National Diploma/Bachelor of Science degree holders were captured. The data distribution shows that 13(31.0%) respondents have had their accounts accessed by unauthorized persons, with the expected count of 13.4(approximately 13). 27(64.3%) of the respondents have not experienced unauthorized access to their accounts. This count would be 22.1 (approximately 22) if no association exists between education qualification and end-user Password authentication experience. 1(2.4%) respondent do not know of such unauthorized access to his account while another 1(2.4%) respondent cannot remember it's occurrence. Expected counts of 3.6(approximately 4) and 2.9(approximately 3) were recorded for option items "Don't Know" and "Can't Remember" respectively.

In the last education qualification category, Ph.D. holders' authentication experiences were captured. Observed counts of 13(28.3%) were recorded for respondents whose accounts have been accessed by unauthorized persons. The expected count where no association exists between education level and end-user Password authentication item would be 14.7(approximately 15). 22 (47.8%) respondents amongst Ph.D. holders have not had their accounts accessed by unauthorized persons. The expected count for this option item is 24.2(approximately 24) assuming no association exists between education level and end-user Password authentication experience item. 4(8.7%) observed counts for Ph.D. holders who do not know of unauthorized access to their accounts, with an expected count of 4, and 7(15.2%) observed counts recorded for respondents who cannot remember if they have had their accounts accessed without authorization against 3.2 (approximately 3) expected counts for this option item.

**Assumption:** The assumption for a table greater than 2 x 2 (i.e., 2 categories by 2 categories) is that not more that 20% of expected counts is less than 5. This applies to our 4 x 4 table.

The closeness between the observed and expected counts in the of education qualification (independent variable) and end-user Password authentication experience (dependent variable) cross tabulation implies that there is little or no association between education level of respondents and their end-user Password authentication experience. Chi-Square tests result is displayed in Table 13 below:

**Table 13**: Chi-Square Tests

|  | Value | Df | Asymp. Sig(2-sided) | Exact Sig. (2 sided) | Exact Sig.(1-sided) |
|---|---|---|---|---|---|
| Pearson Chi-Square | 11.848 [a] | 9 | 0.222 | [b] |  |
| Likelihood Ratio | 12.775 | 9 | 0.173 | 0.222 |  |
| Fisher's ExactTest | 9.909 |  |  | 0.317 |  |
| Linear-by-Linear Association | 1.051 | 1 | 0.305 | [b] | [b] |

| N of Valid Cases | 291 | | | | |
|---|---|---|---|---|---|

a. 6 cells (37.5%) have expected count less than 5. The minimum expected count is 1.10
b. Not computed

From the table above, the assumption is violated since more that hence values will be read from the Likelihood Ratio row.

**Statistical Value:** 12.775; **Degrees of Freedom:** 9; **P-Value:** 0.173

In Table 13, the p-value, which is 0.173, is more than the accepted significance level of 0.05. This implies that there is not enough evidence to conclude that a significant relationship exists between the two variables. Thus, we accept **H0**: Level of education of Internet users have no impact on their end-user experience in Password-based authentication. While the alternative hypothesis: Level of education of Internet users has impact on their end-user experience in Password-based authentication is rejected.

To further ascertain the impact education level has on end-user password authentication experience, Internet users' experience was captured. Table 14 below shows the observed and expected counts for four categories of education qualification cross tabulated with four scale categories of end-user Password authentication experiences, resulting in a 4 x 4 table.

**Assumption:** The assumption for a table greater than 2 x 2 (i.e., 2 categories by 2 categories) is that not more that 20% of expected counts is less than 5. This applies to our 4 x 4 table.

Rating on end-user Password authentication experience was carried out using respondents self-report on their authentication experiences; scaled according to their perceived rating of the experience. Table 14 below displays the crosstabulation result of respondents' gender and their self-rated Password authentication experience.

**Table 14**: Education Qualification * User Experience (How has your experience been using Password authentication?)

| Highest Education Qualification (HEQ) | | How has your experience been using Password Authentication | | | | Total |
|---|---|---|---|---|---|---|
| | | Excellent | Good | Average | Poor | |
| School Certificate | Count | 77 | 92 | 17 | 1 | 187 |
| | Expected Count | 81.0 | 89.3 | 14.8 | 1.9 | 187.0 |
| | % Within HEQ | 41.2% | 49.2% | 9.1%% | 0.5% | 100.0% |
| OND | Count | 8 | 5 | 3 | 0 | 16 |
| | Expected Count | 6.9 | 7.6 | 1.3 | 0.2 | 16.0 |
| | %Within HEQ | 50.0% | 31.2% | 18.8% | 0.0% | 100.0% |
| HND/B.Sc | Count | 19 | 22 | 1 | 0 | 42 |
| | Expected Count | 18.2 | 20.1 | 3.3 | 0.4 | 42.0 |
| | % Within HEQ | 45.2% | 52.4% | 2.4% | 0.0% | 100.0% |
| Ph.D | Count | 22 | 20 | 2 | 2 | 46 |
| | Expected Count | 19.9 | 22.0 | 3.6 | 0.5 | 46.0 |
| | % Within HEQ | 47.8% | 43.5% | 4.3% | 4.3% | 100.0% |
| Total | Count | 126 | 139 | 23 | 3 | 291 |
| | Expected Count | 126.0 | 139.0 | 23.0 | 3.0 | 291.0 |
| | % Within HEQ | 43.3% | 47.8% | 7.9% | 1.0% | 100.0% |

Analysis of the impact of education level of respondents on their Password authentication experiences showed observe counts of 77(41.2%), and an expected count of 81 for respondents who reported having excellent Password authentication experiences in the school certificate holder's category. 92(49.2%) of the respondents in this category have good experiences, with expected count of 89.3(approximately 89). Respondent with average Password authentication experiences had a distribution of 17(9.1%) observed counts and 14.8(approximately 15) expected counts, while those with poor Password authentication experiences had 1(0.5%) observed and 1.9 (approximately 2) expected counts.

The second category captured respondents with National Diploma. Distribution in this category show that 8(50.0%) observed counts for respondents that have excellent Password authentication, with an expected count of 6.9(approximately 7); 5(31.2%) observed counts and 7.6 (approximately 8) expected counts for respondents that reported having good Password authentication experiences; 3(18.8%) observed counts and 1.3(approximately 1) expected count for respondents that reported having average Password authentication experienced.

Lastly, no respondent reported having poor password authentication experience, with an expected count of 0.2(approximately 0), also, was reported. Following the National Diploma category are respondents whose highest qualification is Higher National Diploma or bachelor's degree. 19(45.2%) observed counts and 18.2(approximately 18) expected counts for respondents that reported having excellent Password authentication experiences; 22(52.4%) observed counts and 20.4(approximately 20) expected counts for respondents who reported having good password authentication experiences; 1(2.4%) observed counts and 3.3(approximately 3) expected counts for respondents who have average password authentication experiences; no observed count and 0.4(approximately 0) expected count for respondents who reported having poor Password authentication experiences.

The fourth category captured respondents with Ph.D. as highest education qualification. Distribution in this category showed 22(47.8%) observed counts and 19.9(approximately 20) expected counts for respondents who reported having excellent password authentication experiences; 20(43.5%) observed counts and 22 expected counts for respondents with good password authentication experiences; 2(4.3%) observed counts and 3.6(approximately 4) expected counts for respondents with average password authentication experiences, and lastly, 2(4.3%) observed counts and 0.5(approximately 1) expected count for poor password authentication experience.

The disparity between observed and expected counts indicate that an association exists between the independent variable (education qualification) and dependent variable (end-user password authentication experience). The strength of this association is further tested using Chi-Square tests.

For a 4 x 4 (4 categories by 4 categories) table such as the Table 14, Chi-Square test also shows if the assumption of having not more than 20% of cells containing expected counts of less than 5. The result of Chi-Square test is displayed in Table 15 below:

**Table 15**: Chi-Square Tests

|  | Value | Df | Asymp. Sig(2-sided) | Exact Sig. (2 sided) | Exact Sig.(1-sided) | Point Probability |
|---|---|---|---|---|---|---|
| Pearson Chi-Square | 12.998 [a] | 9 | 0.163 | [b] |  |  |
| Likelihood Ratio | 11.683 | 9 | 0.232 | 0.236 |  |  |
| Fisher's ExactTest | 11.037 |  |  | 0.216 |  |  |
| Linear-by-Linear Association | 0.74 [c] | 1 | 0.541 | 0.555 | 0.283 | 0.020 |
| N of Valid Cases | 291 |  |  |  |  |  |

a. **7 cells (43.8%) have expected count less than 5. The minimum expected count is 0.16**
b. **Not computed**
c. **The standardized statistic is -0.612**

From Table 15 above, 43.8% (7 cells) have expected counts of less than 5. This violates the assumption that not more than 20% of cells should have expected counts less than 5. The value readings are therefore taken from the Likelihood Ratio row.

**Statistical Value:** 11.683; **Degree of Freedom:** 9; **P-Value:** 0.232

In Table 15, the reported p-value is 0.232. This p-value is greater than 0.05 (the acceptable significant level). This means that the test results are not statistically significant at 0.05 level. In other words, we cannot reject the null hypothesis (**H0**) that there is no effect or relationship between the variables being tested. We therefore accept the Null Hypothesis(**H0**): Level of education of Internet users has no impact on their end-user experience in Password-based authentication and reject the alternative hypothesis(**H1**): Level of education of Internet users has impact on their end-user experience in Password-based authentication.

It is important to note that when conducting a hypothesis test, the null hypothesis is typically rejected if the p-value is less than the significance level. The significance level is the probability of rejecting the null hypothesis when it is true. The most used significance level is 0.05, which means that there is 5% chance of rejecting the null hypothesis when it is true.

## 3.3 Discussion

The results of our findings showed that password experience of end-users in these academic communities remain fair and good. An indication that the password-based experiences of end-users in developed and highly technological advanced economies did not reflect the experiences of end-users in this geographical area, and academic communities. The study also aligned the experiences with factors such as age (under 60), highest educational qualification, and employment status (unemployed, employed as temporary or permanent staff), but the result failed to establish a relationship with these factors. Related studies by Ugwu *et al*., (2022) and Ugwu *et al*., (2023) conducted earlier within the same geographical areas and academic communities suggest that age, gender, educational qualification, and academic discipline of the respondents have no effect on their personal cyber hygiene culture in ICT usage. These studies when juxtaposed with the current study, provided an insight that perhaps, (1) that there is likelihood of a relationship between cyber hygiene culture and password use experiences, and (2) that these experiences are not dependent on age, gender, highest educational qualification, and academic discipline while in the University. All the same, this study did not focus on understanding the password experiences of very economic buoyant communities as there are relationship between possibilities of cyber-attacks on very economic viable communities. This statement is supported by a report by Mottl, (2022), which suggests that cyber-attack incidents are highly motivated by economic gains. The above informs why ICT end-users are very conscious of having strong password protection for their financial accounts as having password leakage or compromise could have the most irreparable damage on end-users (Ray, H., *et al*., 2021).

According to Aladenusi, T. and Odumuboni, F. (2022), Nigeria's cyber security outlook for 2022 showed a rise in cyber-attacks and data breach incidents both in public, private financial and non-financial sectors. An affirmation that no sector of the economy is spared of this menace with a prediction that Nigeria has come under the spotlight of cyber-attackers armed with more tools and ground to play. Password compromise is one of the targets of cyber-criminals and major motivation for this remains economic. This information alongside existing knowledge gaps stated earlier deepened the passion of the researcher in surveying password experience in south-eastern Nigeria with particular interest on the staff and students at some University communities therein to understand their password use experiences.

Password experience the authors argue appears to be related to some factors ranging from cyber hygiene, economic status, political environment and so on when the outcome of this study is compared with other studies conducted in different economic zone and climes. For instance, Beyond Identity (2021) reveals that 46% of US consumers failed to complete transactions due to authentication failure. Lance (2021) shows that 67% of the respondents of their study conducted in the USA said that they forgot their password when they are trying to complete an online banking transaction, 56% said it happens when trying to get travel information, 55% reported it happens when they are attempting to buy something, and 43% said it happens when they try to access a document.

The works of Chiasson and Biddle, (2007); Butler and Butler, (2015) are of the views that password-based authentication are problematic. Butler and Butler (2015) opined that because of unlimited online options, end-users have become impatient with the time-consuming and inconvenient login experiences. These experiences include being prompted to reset a Password or create an account with a long-form and managing Passwords (measures related to the safekeeping of Passwords). These experiences are expressed by social media users in a study conducted by Majid and Kouser (2019), which reported that many of their respondents that are social media users forgot to log out of their social media accounts leading to possible account compromise especially in public computers.

Furthermore, Beyond Identity (2021) reveals that 46% of US consumers failed to complete transactions due to authentication failure. Also, 18.75% of returning users abandon the cart after forgetting their password and having issues with password reset emails. Likewise, 36% of the consumers said they try to guess a forgotten password twice before resetting it, while 28% said they guess a forgotten password once and 22% three times. Singh *et al.,* (2007) describe users' experiences of how an entire village would delegate bank credentials to a single person who would conduct the online transaction on everybody's behalf. In addition, Dunphy *et al*., (2014: 2) report on how older adults delegate the withdrawal of their monies from an automated teller machine to their helpers thereby disclosing their personal identification numbers (PINs) to a third party. The same experience is also shared by electronic health clients, where the issues of forgetting passwords, password reset and so on could be critical, dangerous or fata (Stantinides *et al.,* (2020)).

Password experience of end-users from this and related studies appears to be hugely relational and could be based on individual geographical location, economic and societal status, personal cyber security consciousness, personal cyber hygiene culture, password practices, and so on.

## 4.0     Recommendations
Based on the above findings, the following recommendations are suggested to further enhance the password hygiene for end-users of password-based authentication systems.

## 4.1     Recommendations for end-users
**Periodic change of password**: Periodic change of password is essential for effective security. This is because most password-based hacks have more to do with bad password, phishing attacks, and keystroke loggers (Isobe and Ito, 2021). Also, with the emerging trends of technology, the possibilities of new computing devices may emerge, and the pattern of usage may change (Ofusori, 2019). Hence, periodically changing your password helps avoids common threats.

**Continuous user education and security awareness:** It is the responsibility of end-users to ensure that they are updated with the latest scientific knowledge on password security best practices by undertaking more training on digital literacy programs and security awareness. This continuous education can help foster a culture of responsible password usage such as creating strong passwords, avoiding common password mistakes, and understanding the importance of password hygiene. Furthermore, having this knowledge will help them avoid the risks associated with password-based authentication system (Renaud, Otondo and Warkentin, 2019)

**Two-Factor Authentication (2FA):** As an additional layer of security, it is recommended that two-factor authentication method should be adopted. This can involve combining passwords with another authentication factor, such as SMS-based OTPs, app-based OTPs, or biometrics when feasible. While it is noted that single authentication such as password authentication is popularly in use as a security measure, two-factor authentications has been proven to be a better effective security measure, reducing the likelihood of unauthorized access fraud (Lambiase *et al.*, 2020)

**Password recovery mechanism:** Password recovery mechanisms are important for users who may forget their passwords. In developing economies, where access to email or mobile services may be limited, alternative method for password recovery, such as security questions or offline recovery options can be used to regain access to their accounts (Majid and Kouser, 2019).

**Password Managers**: It is suggested that end-users are educated on the benefits of using password managers to simplify and enhance password management.  The use of password manager applications will help users generate and securely store complex passwords (Fagan *et al*., (2017). Furthermore, it helps users avoid the pitfall of reusing passwords and simplify the process of managing multiple passwords

**Continuous Monitoring and Updates**: Regularly monitor authentication systems for security vulnerabilities and apply timely updates and patches. It is important to note that keeping computing devices, operating systems and applications up to date with the latest security patches, helps ensure that user accounts are protected against emerging threats and security breaches. Outdated software can have vulnerabilities that attackers can exploit to gain unauthorized access to passwords (AlSabah, Oligeri and Riley, 2018).

## 4.2 Recommendations for policy makers

**Password Policies**: Policymakers in developing economies can implement simplified password policies that strike a balance between security and usability. This can be done by avoiding overly complex requirements that may lead to users creating weak or easily forgotten passwords. For example, encourage longer passphrases instead of traditional passwords (Güven, Boyaci and Aydin, 2022).

**Collaborative Efforts:** Policymakers in developing economies can foster collaboration between governments, private organizations, and technology providers to address the challenges of password-based authentication. Pool resources and expertise to develop localized solutions, share best practices, and support initiatives aimed at improving user experiences and security.

By implementing these recommendations, the usability and security of password-based authentication can be enhanced, leading to better experiences for end users in developing economies.

## 5. Research contributions and suggestions for future research

This study provides insights into end-users' password-based experience in developing economies to ascertain whether their experience is associated with educational qualification, age and gender. Furthermore, it provides an empirical basis on which future studies on password-based authentication and end-user experiences can be built upon. Also, the outcome of this research will help inform policymakers and research communities towards password hygiene culture, management, and the feasibility of passwordless authentication systems in developing economies.

It is important to note that this study did not delve into understanding the relationship between password experience and the economic and social status of individuals. Hence, future work should incorporate this area by having a mix of people from different societal and economic levels. In addition, further research can be done to investigate the relationship between ppassword entropy and end-users experience towards maintaining password hygiene.

## 6. Conclusion

This work examined the experience of end-users of password-based authentication using questionnaire. The survey followed an established model used in information system in studying users' experiences in adopting new technologies. The result showed that the issues of account compromise in the geographical area is not common as the respondents reported having good experience with password-based authentication systems. Furthermore, that both age and educational qualifications have no impact on end-user experience as shown in this study. In all, the age of the respondents is under 60 and the result of this study may not apply to ages over 60 that may have problems with remembering passwords, people with multiple accounts that do not use Password management utility such as Password Managers.

All the same, because this study did not consider a measure of end-users Password entropy, use of Password managers and Password hygiene culture, the results may also vary when these are put into consideration. This study concludes that password experience is relational and varies in geographical locations, communities, economic and societal status, ICT tool usage and engagements. On the basis of these, the authors recommend that end-users of ICT tools should endeavour to take the security of their password management serious as that largely depends on the level of their cyber hygiene culture, consciousness, and discipline.

Di Leo G and Sardanelli F (2020) Statistical significance: p value, 0.05 threshold, and applications to radiomics—reasons for a conservative approach. *European radiology experimental* 4(1): 1-8.

Dunphy, P., Monk, A., Vines, J., Blythe, M., & Olivier, P. (2014). Designing for spontaneous and secure delegation in digital payments. Interacting with Computers, 26(5), 417-432.

Dunphy, P., Vlachokyriakos, V., Thieme, A., Nicholson, J., McCarthy, J., & Olivier, P. (2015, July). Social media as a resource for understanding security experiences: A qualitative analysis of# password tweets. In Symposium on Usable Privacy and Security (SOUPS).

Fagan M, Albayram Y, Khan MMH, *et al*. (2017) An investigation into users' considerations towards using Password managers. *Human-centric Computing and Information Sciences* 7(1): 1-20.

Furnell, S. (2022). Assessing website password practices–Unchanged after fifteen years?. Computers & Security, 120, 102790.

Güven, E. Y., Boyaci, A., & Aydin, M. A. (2022). A novel password policy focusing on altering user password selection habits: a statistical analysis on breached data. Computers & Security, 113, 102560.

Han, S., Skinner, G., Potdar, V., & Chang, E. (2006, November). A framework of authentication and authorization for e-health services. In Proceedings of the 3rd ACM workshop on Secure web services (pp. 105-106).

Huerta-Álvarez R, Cambra-Fierro JJ and Fuentes-Blasco M (2020) The interplay between social media communication, brand equity and brand engagement in tourist destinations: An analysis in an emerging economy. *Journal of Destination Marketing & Management* 16: 100413.

Isobe T and Ito R (2021) Security analysis of end-to-end encryption for zoom meetings. *IEEE access* 9: 90677-90689.

Jakkal, V. (2021). The passwordless future is here for your Microsoft account. [Online] https://www.microsoft.com/en-us/security/blog/2021/09/15/the-passwordless-future-is-here-for-your-microsoft-account/. Accessed: 17/05/2023

Krejcie, R. V., & Morgan, D. W. (1970). Determining sample size for research activities. Educational and psychological measurement, 30(3), 607-610.

Krol, K., Philippou, E., De Cristofaro, E., & Sasse, M. A. (2015). "They brought in the horrible key ring thing!" Analysing the Usability of Two-Factor Authentication in UK Online Banking. *arXiv preprint arXiv:1501.04434*.

Lambiase, M. V., Grajek, G. F., Jeffery Chiwai, LO., & Tommy Ching Hsiang, WU. (2020). Single sign on with multiple authentication factors. Google Patents.

Lance W (2021) *How Password troubles could cost your online business potential sales*. Available at: https://www.techrepublic.com/article/how-Password-troubles-could-cost-your-online-business-potential-sales/ (accessed 11 May).

Leedy, P. & Orrnrod, J. (2005). *Practical research: Planning and design.* New Jersey: Prentice Hall.

Liu, Y., Zhang, L., Yang, Y., Zhou, L., Ren, L., Wang, F., Liu, R., Pang, Z. and Deen, M.J., (2019). A novel cloud-based framework for the elderly healthcare services using digital twin. IEEE access, 7, pp.49088-49101.

Majid I and Kouser S (2019) Social Media and Security: How to Ensure Safe Social Networking. *Majid, I & Kouser, S.(2019). Social media and security: how to ensure safe social networking. International Journal of Humanities and Education Research* 1(1): 36-38.

Meter DJ and Bauman S (2015) When sharing is a bad idea: The effects of online social network engagement and sharing Passwords with friends on cyberbullying involvement. *Cyberpsychology, Behavior, and Social Networking* 18(8): 437-442.

Morrison B, Coventry L and Briggs P (2021) How do Older Adults feel about engaging with Cyber-Security? *Human Behavior and Emerging Technologies* 3(5): 1033-1049.
29